\documentclass[12pt,english]{article}
 \usepackage{helvet}
 \usepackage[T1]{fontenc}
 \usepackage[cp1251]{inputenc}
 \usepackage{geometry}
 \usepackage{graphicx}
 \usepackage{setspace}
 \usepackage{babel}

 \renewcommand{\vec}[1]{\mbox{\boldmath $#1$}}

 \def\gsim{\lower.4ex\hbox{$\;\buildrel >\over{\scriptstyle\sim}\;$}}
 \def\lsim{\lower.4ex\hbox{$\;\buildrel <\over{\scriptstyle\sim}\;$}}
 
 \def\bl{\par\vskip 12pt}
 \def\bll{\par\vskip 24pt}
 \def\blll{\par\vskip 36pt}

 \def\beg{\begin{eqnarray}}
 \def\ende{\end{eqnarray}}

 \def\n{${\cal N}^{\underline\circ}$}

\begin{document}

\begin{center}
{\bf THE SOLAR DYNAMO: INFERENCES FROM OBSERVATIONS} \\[0.2 truecm]
{\bf AND MODELING}
\end{center}

\bll

\centerline{L.\,L.~Kitchatinov$^{1,2}$}

\bl

\begin{center}
$^1${\it Institute for Solar–Terrestrial Physics, P.O. Box 4026, Irkutsk, 664033 Russia \\ e-mail: kit@iszf.irk.ru} \\
$^2${\it Pulkovo Astronomical Observatory, Pulkovskoe Sh. 65, St. Petersburg, 196140 Russia}
\end{center}

\bll
\hspace{0.8 truecm}
\parbox{14.4 truecm}{
{\bf Abstract} -- It can be shown on observational grounds that two basic effects of dynamo theory for solar activity – production of the toroidal field from the poloidal one by differential rotation and reverse conversion of the toroidal field to the poloidal configuration by helical motions – are operating in the Sun. These two effects, however, do not suffice for constructing a realistic model for the solar dynamo. Only when a non-local version of the $\alpha$-effect is applied, is downward diamagnetic pumping included and field advection by the equatorward meridional flow near the base of the convection zone allowed for, can the observed activity cycles be closely reproduced. Fluctuations in the $\alpha$-effect can be estimated from sunspot data. Dynamo models with fluctuating parameters reproduce irregularities of solar cycles including the grand activity minima. The physics of parametric excitation of irregularities remains, however, to be understood.
 }

\bll


\blll

\reversemarginpar

\setlength{\baselineskip}{0.8 truecm}

%%%%%%%%%%%%%%%%%%%%%%%%%%%%%%%%%%%%%%%%%%%%%%%%%%%%%%%%%%%%%%%%%%%
 \bll
 {\bf 1. INTRODUCTION}
 \bl
%%%%%%%%%%%%%%%%%%%%%%%%%%%%%%%%%%%%%%%%%%%%%%%%%%%%%%%%%%%%%%%%%%%

The magnetic activity of the sun controls space weather conditions in the heliosphere and influences the Earth. A well-known example of this influence is the geomagnetic storm of 13 March 1989, which caused a breakdown in electric power supply for about 9 hrs in Quebeck,  Canada. The powerful solar flare that caused this storm had energy of the order of 10$^{32}$ erg (Benz 2008). Statistics of the Kepler mission show that flares of 100 times larger energy occur on sun-like stars with an average frequency of one event in about 800 yrs (Maehara et al. 2012). Such powerful flares – if they occur on the Sun (Shibata et al. 2013) – present a real concern for modern civilization.

The process behind solar magnetic activity is believed to be a hydromagnetic dynamo operation in the solar interior. The basic ideas of the solar dynamo theory were formulated long ago (Parker 1955; Babcock 1961; Steenbeck et al. 1971) and already the first numerical models based on the theory produced oscillating solutions reminiscent of the solar cycle (Leighton 1969; Steenbeck \& Krause 1969; K\"ohler 1973; Ivanova \& Ruzmaikin 1976). Considerable discrepancies between dynamo models and solar observations were, however, present and persisted for decades. Progress was inhibited by the absence of observational information on the internal dynamo process. It may be noted in this relation that differential rotation theory guided by helioseismology seems to be developing much faster, though the theories of global flows and global magnetic fields use essentially the same methods. The last decade has changed the situation by providing observations-based inferences on the main dynamo effects.

This review focuses on observational hints about solar dynamo mechanisms and on recent numerical models of the solar dynamo that started reproducing solar activity quite closely having incorporated observational inferences. The next section discusses the main dynamo effects and what observations tell us about their operation on the Sun. Section 3 discusses difficulties of solar dynamo modeling and how contemporary numerical models attempt to resolve the problems. The final section 4 concludes by summarizing recent progress and possibilities for further developments.

%%%%%%%%%%%%%%%%%%%%%%%%%%%%%%%%%%%%%%%%%%%%%%%%%%%%%%%%%%%%%%%%%%%
 \bll
 {\bf 2. TWO BASIC EFFECTS: HINTS FROM OBSERVATIONS}
%%%%%%%%%%%%%%%%%%%%%%%%%%%%%%%%%%%%%%%%%%%%%%%%%%%%%%%%%%%%%%%%%%%

 {\bf 2.1. Solar Cycle Scenario}
 \bl
%%%%%%%%%%%%%%%%%%%%%%%%%%%%%%%%%%%%%%%%%%%%%%%%%%%%%%%%%%%%%%%%%%%

Since the pioneering work of Parker (1955), the solar dynamo is believed to be driven by two main processes: the winding of the global (e.g., longitude-averaged) toroidal field from the poloidal one by differential rotation and conversion of the toroidal field back to the poloidal configuration by helical motions. These two processes named the $\Omega$- and $\alpha$-effects, respectively, lead to the solar cycle scenario shown in Fig. 1.

\begin{figure}[htb]
 \centerline{
 \includegraphics[width=14 cm]{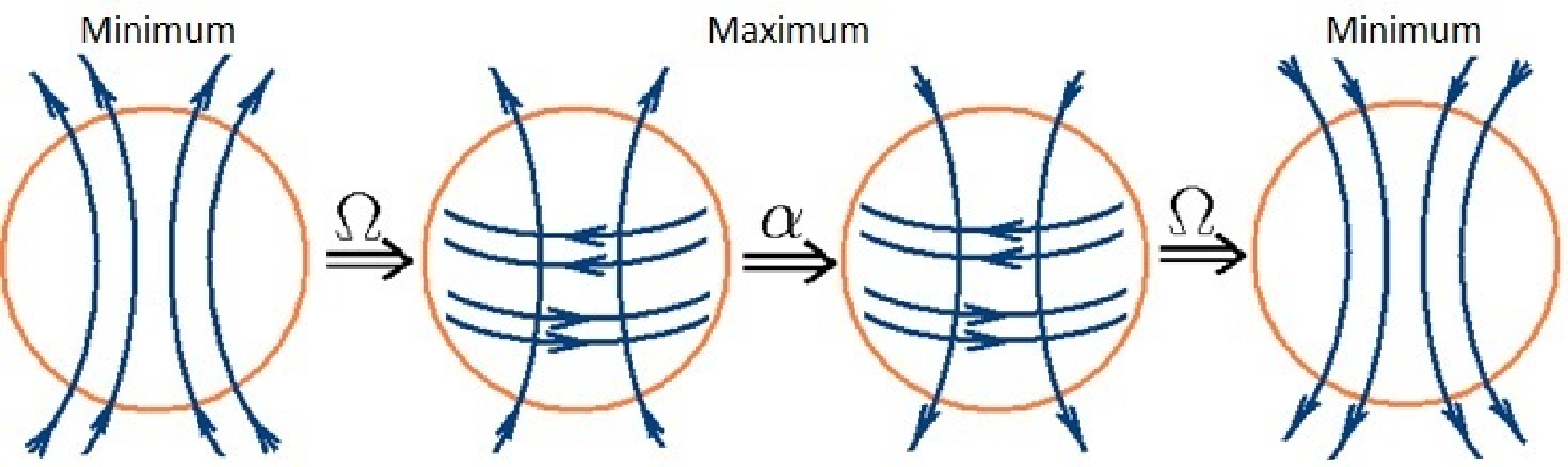}}
 \begin{description}
 \item{\small Fig.~1. Solar cycle scenario based on the $\alpha\Omega$-dynamo mechanism.
    }
 \end{description}
\end{figure}

The left panel of Fig.1 illustrates the minimum activity phase of the solar cycle. This phase is dominated by the relatively weak (1$\div$2 G) poloidal field. The differential rotation winds the toroidal field, which has opposite signs in the northern and southern hemispheres. From this toroidal field, the alpha effect produces a new poloidal field opposite in sign to the original poloidal field. The poloidal field decreases, but the differential rotation still amplifies the toroidal field until the poloidal field reverses. After the reversal, the differential rotation starts producing a new toroidal field opposite in sign to the present one, and the toroidal field starts decreasing. While it decreases, the alpha effect still amplifies the reversed poloidal field until the toroidal field becomes small at the epoch of a new activity minimum.

Sunspots are associated with the emergence of toroidal flux tubes on the surface from the deep solar interior. Spots of bipolar sunspot groups, leading in rotational motions, typically have opposite magnetic polarities in the northern and southern hemispheres in accord with Fig.1. The polar field reverses at the maximum phase of a sunspot cycle, and the global magnetic field has opposite signs in successive solar cycles. The scenario in Fig.1, therefore, qualitatively agrees with observations. Quantitative agreement for numerical dynamo models is, however, much more problematic.

%%%%%%%%%%%%%%%%%%%%%%%%%%%%%%%%%%%%%%%%%%%%%%%%%%%%%%%%%%%%%%%%%%%
 \bll
 {\bf 2.2. The $\Omega$-Effect}
 \bl
%%%%%%%%%%%%%%%%%%%%%%%%%%%%%%%%%%%%%%%%%%%%%%%%%%%%%%%%%%%%%%%%%%%

Schatten et al. (1978) were probably the first to notice that the scenario in Fig.1 implies a functional relation between poloidal field amplitude at the activity minimum and the strength of the following activity cycle. This is because the solar differential rotation is very regular. It varies little with time. The relative amplitude of torsional oscillations (Howard \& LaBonte 1980; Vorontsov et al. 2002) is small. The $\Omega$-effect producing a toroidal field from a poloidal one is, therefore, expected to provide the aforementioned functional relation. Makarov \& Tlatov (2000) and Makarov et al. (2001) have shown that solar cycles 16 to 22 indeed closely followed a linear relation between the cycles’ strengths and the poloidal field at the preceding activity minima.

\begin{figure}[htb]
 \centerline{
 \includegraphics[width=12 cm]{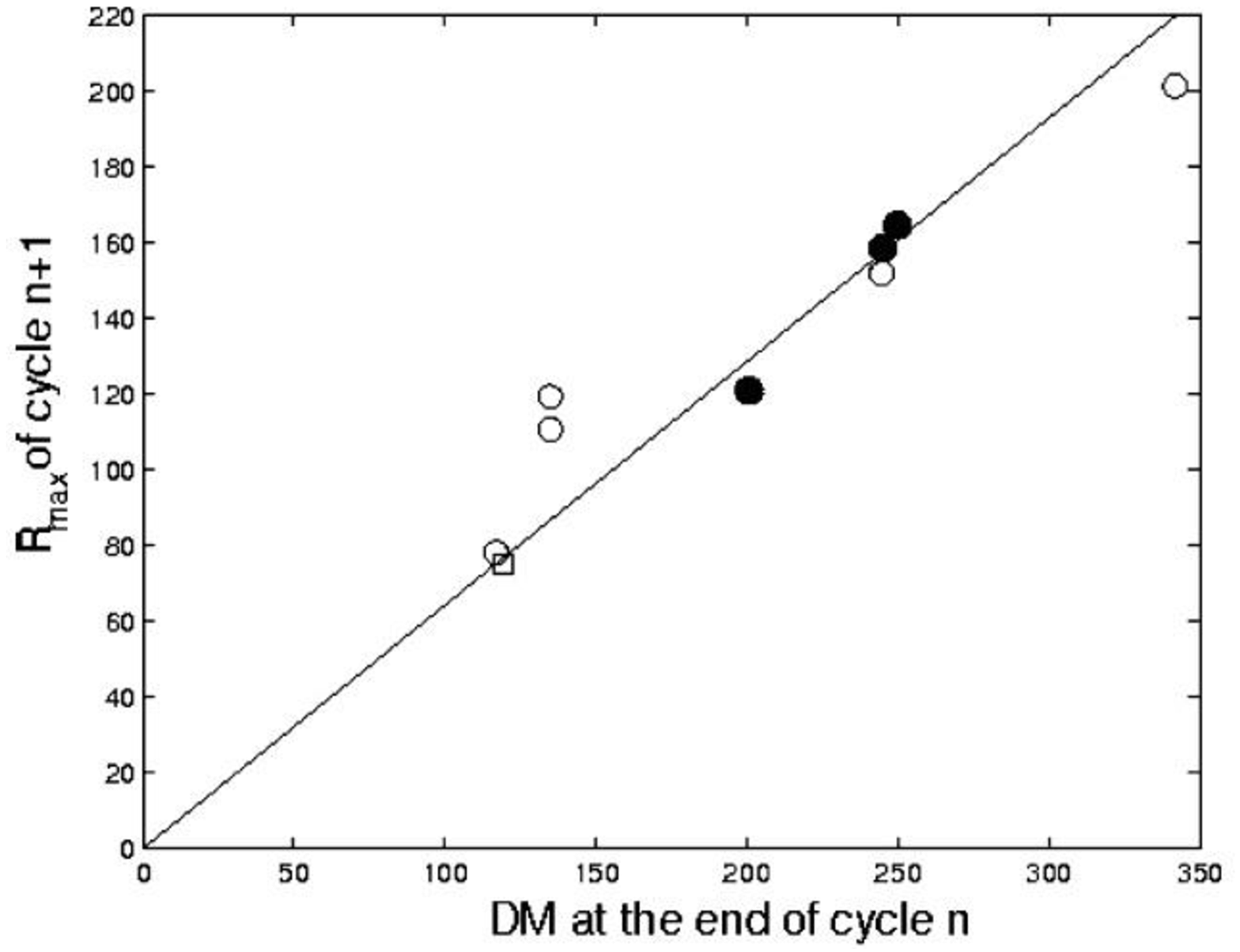}}
 \begin{description}
 \item{\small Fig.~2. Maximum strength of sunspot cycles R$_\mathrm{max}$ as a function of the dipole momentum (DM) of the polar magnetic field of the preceding solar minima. Filled and open circles stand for DM measurements and DM estimations from the A-index of the large-scale field, respectively (see text). The square shows the strength of the 24th cycle predicted from the linear relationship of this plot. From Jiang et al. (2007).
    }
 \end{description}
\end{figure}

Figure 2 shows the positions of the last 9 solar cycles, including the current 24th cycle, on the plane of the cycle strength R$_\mathrm{max}$  and dipole momentum DM of the polar field (Svalgaard et al. 2005) of the preceding solar minimum (R$_\mathrm{max}$ is the  maximum value of the annual running mean of the  international sunspot number in a given cycle). The open circles in the Figure stand for the DM estimated by rescaling the A-index of the large-scale field of Makarov \& Tlatov (2000) (cf. Jiang et al. (2007) for the rescaling procedure). The filled circles show direct measurements of the polar magnetic field. The observations-based DM values are available for the four last solar minima (Svalgaard et al. 2005). Only three filled circles in Fig.2 mean that the strength of the current 24th cycle was not known at the time this Figure was constructed by Jiang et al. (2007). The position of the 24th cycle was shown by the square in the plot as a prediction of the current cycle strength. Judging from the current level of solar activity, the prediction is quite accurate. The close correlation of Fig.2 leaves little doubt about the $\Omega$-effect’s participation in the solar dynamo.

%%%%%%%%%%%%%%%%%%%%%%%%%%%%%%%%%%%%%%%%%%%%%%%%%%%%%%%%%%%%%%%%%%%
 \bll
 {\bf 2.3. The $\alpha$-Effect}
 \bl
%%%%%%%%%%%%%%%%%%%%%%%%%%%%%%%%%%%%%%%%%%%%%%%%%%%%%%%%%%%%%%%%%%%

As the $\alpha$-effect converts the toroidal field into a poloidal one (Fig.1), a functional relation may naively be expected between amplitudes of solar cycles and poloidal fields of following minima, similar to the tight correlation of Fig.2. The cycles’ strengths R$_\mathrm{max}$, however, do not correlate with DM at the cycles’ ends (cf., Fig.3 of Jiang et al. (2007)). In contrast with the regular $\Omega$-effect, considerable randomness is inherent to the $\alpha$-effect. The $\alpha$-effect is related to relatively small-scale motions, which are not regular in contrast to global differential rotation.

In the main, two types of $\alpha$-effect are discussed in literature: the effect of inhomogeneous convective turbulence originally proposed by Parker (1955) and the Babcock-Leighton mechanism associated with surface active regions (Babcock 1961). In both cases, the $\alpha$-effect results from the influence of the Coriolis force on associated motions but the motions are of a different nature: thermal convection in the case of a "classical" $\alpha$-effect by Parker (1955) and magnetic buoyancy in the case of the Babcock-Leighton mechanism. The theory of the $\alpha$-effect is mainly focused on its classical version where standard tools of quasi-linear theory can be applied. The Babcock-Leighton mechanism deserves, however, not less attention because its properties can be estimated from sunspot statistics and there are indications of its predominance on the Sun, which are discussed later.

The $\alpha$-effect of Babcock-Leighton type is associated with Joy’s law stating that the spots of bipolar sunspot groups leading in rotational motion are on average closer to the equator than the following spots. The average tilt angle between the line connecting centers of gravity of opposite polarities and the line of constant latitude (Fig.3) is positive and increases with latitude. The magnetic fields of the loops connecting spots of opposite polarities have, therefore, poloidal components which contribute to the global poloidal field when the active regions decay. Distributions of the tilt angles are centered about positive values but are quite broad and extend far in the region of negative values (cf. Fig.11 in Howard (1996)). Random fluctuations are, therefore, inherent to the Babcock-Leighton $\alpha$-effect. The randomness does not, however, prevent estimations of the contributions of this effect to the global poloidal field.

\begin{figure}[htb]
 \centerline{
 \includegraphics[width=12 cm]{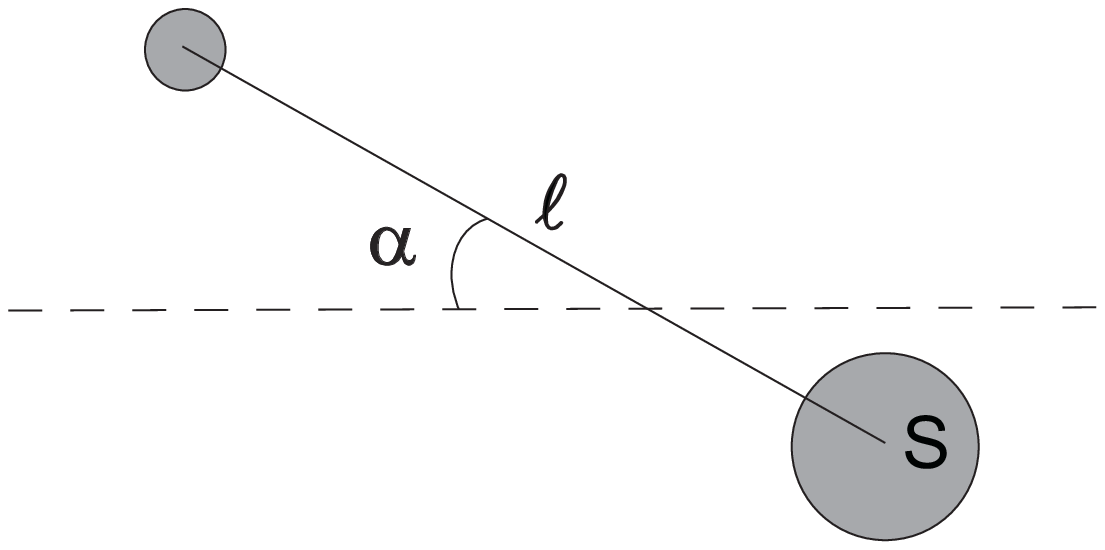}}
 \begin{description}
 \item{\small Fig.~3. Illustration of Joy’s law for solar bipolar spot groups and parameters of Eq.\,(\ref{1}). The dashed line shows the local solar parallel in the northern hemisphere. The equator is downward. The tilt angle $\alpha$ is positive.
    }
 \end{description}
\end{figure}

The estimations using sunspot statistics were first attempted by Erofeev (2004) who found the estimated contributions of the Babcock-Leighton mechanism into the global poloidal field for eight solar cycles to correlate with the poloidal field amplitude of the following activity minima.  Dasi-Espuig et al. (2010) found significant correlation between the product of the cycle strength with the averaged tilt angle for the same cycle and the strength of the next cycle. The contribution of the Babcock-Leighton mechanism to the poloidal field can be estimated by the sum,
\begin{equation}
    B = \sum\limits_{i}^{} S_i \ell_i \sin\alpha_i ,
    \label{1}
\end{equation}
where summation is over sunspot groups, $S$ is the area of the largest spot in the group, $\ell$ is the distance between the centers of gravity of opposite magnetic polarities, and $\alpha$ is the tilt angle (Fig.3); the parameter values are taken for the epoch of maximum development of the group (Kitchatinov \& Olemskoy 2011a).  The value of B$_\mathrm{cyc}$, resulting from the summation in Eq.(1) over the entire activity cycle, correlates well with the A-index of the polar field of Makarov \& Tlatov (2000) of the following activity minimum. The correlation is shown in Fig.4. The B$_\mathrm{cyc}$ values of this plot were estimated using the data of the Catalogue of Solar Activity (CSA) of Pulkovo Astronomical Observatory (Nagovitsyn et al. 2008) as well as sunspot data of Mount Wilson (MW) and Kodaikanal (KK) Observatories digitized by  Howard et al. (1984, 1999). The tight correlation of Fig.4 shows that the Babcock—Leighton $\alpha$-effect, by all probabilities, takes part in the solar dynamo.

\begin{figure}[htb]
 \centerline{
 \includegraphics[width=12 cm]{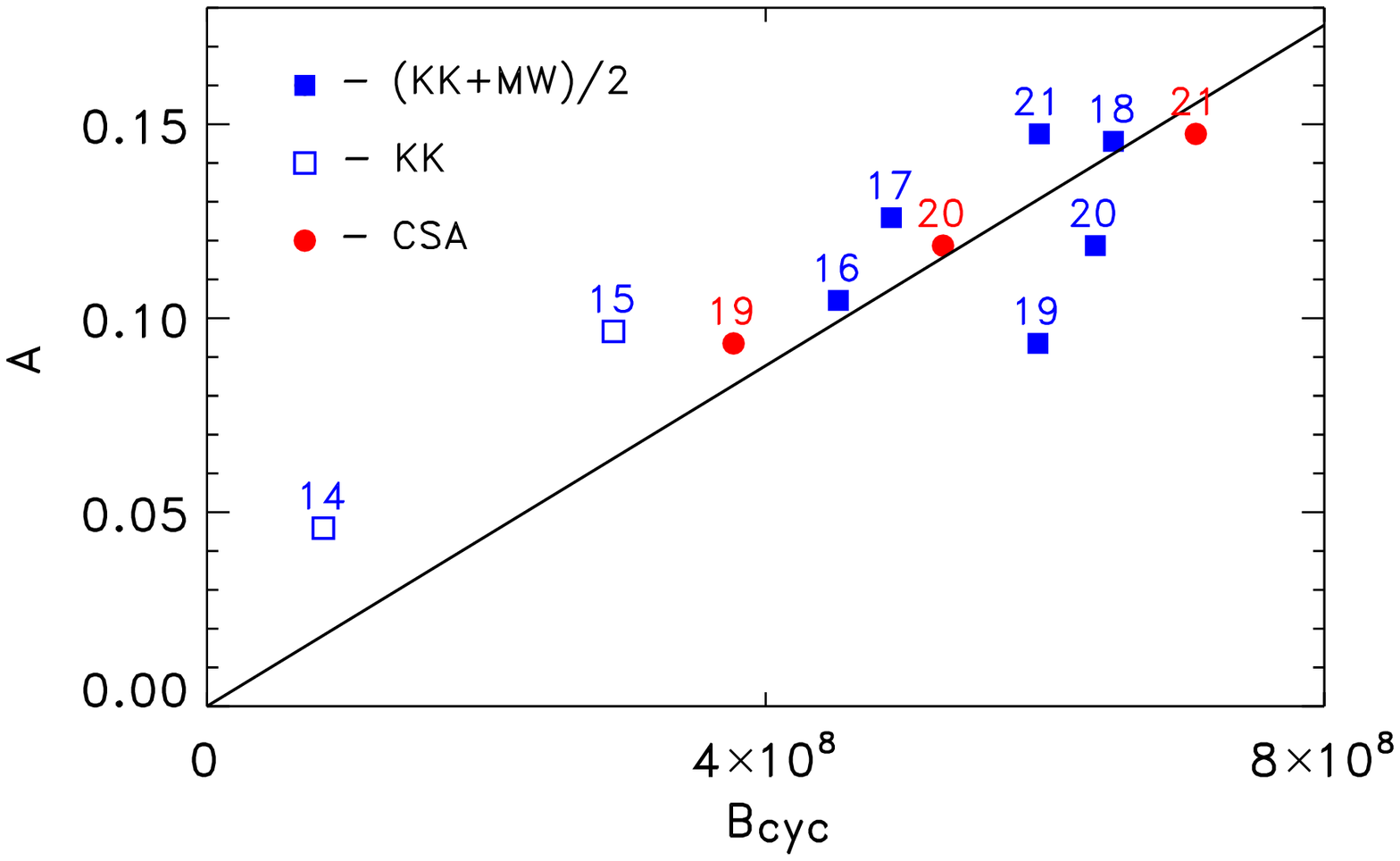}}
 \begin{description}
 \item{\small Fig.~4. Positions of individual solar cycles on the plane of their B$_\mathrm{cyc}$ values and the A-index of the large-scale magnetic field for the activity minima following these cycles. The positions are marked by the solar cycle numbers.  B$_\mathrm{cyc}$ values were estimated with Eq.(\ref{1})  using the data of the Pulkovo Catalogue (CSA) and sunspot data of Mount Wilson (MW) and Kodaikanal (KK) observatories. The strait line shows linear regression. After Olemskoy et al. (2013).
    }
 \end{description}
\end{figure}

Even during the maximum activity epochs, a small number of active regions are simultaneously present on the Sun. Accordingly, the values of $B_\mathrm{Carr}$ estimated with Eq.(1) for individual Carrington rotations fluctuate strongly. Their running mean $\langle B_\mathrm{Carr}\rangle$  over 13 rotations (about one year) vary, however, smoothly with time. The relative value of the fluctuations,
\begin{equation}
    \sigma = \sqrt{\overline{\left( \frac{B_\mathrm{Carr}}{\langle B_\mathrm{Carr}\rangle} - 1\right)^2}} ,
    \label{2}
\end{equation}
where the overline means averaging over the entire dataset of a catalogue, was estimated from CSA data to be $\sigma = 2.7$ (Olemskoy et al. 2013). The Babcock-Leighton $\alpha$-effect shows strong irregular variations. Fluctuations in basic parameters are important for a dynamo. The fluctuations are believed to be the reason for the variable strength of activity cycles and even for the grand minima of solar activity (Hoyng 1988; Choudhuri 1992; Ossendriver et al. 1996; Moss et al. 2008). Incorporation of the fluctuations with the amplitude (\ref{2}) estimated from sunspot data into a dynamo model helps to reproduce the statistics of solar grand minima quite closely (Olemskoy \& Kitchatinov 2013).

%%%%%%%%%%%%%%%%%%%%%%%%%%%%%%%%%%%%%%%%%%%%%%%%%%%%%%%%%%%%%%%%%%%
 \bll
 {\bf 3. SOLAR DYNAMO MODELING}
 \bl
 {\bf 3.1. New Solutions for Old Problems}
 \bll
%%%%%%%%%%%%%%%%%%%%%%%%%%%%%%%%%%%%%%%%%%%%%%%%%%%%%%%%%%%%%%%%%%%

The presence of the two basic dynamo effects discussed in the preceding section on the Sun is supported by observations. It is not surprising, therefore, that already the first numerical models based on these effects (cf., e.g., Steenbeck \& Krause 1969; K\"ohler 1973; Stix 1976) gave oscillating solutions similar to the solar cycle.  Large quantitative discrepancies with observations were, however, present and persisted for decades. The situation has largely improved since then. We proceed by discussing several effects of crucial significance for solar dynamo modeling.
%%%%%%%%%%%%%%%%%%%%%%%%%%%%%%%%%%%%%%%%%%%%%%%%%%%%%%%%%%%%%%%%%%%
 \bll
 {\bf 3.1.1. Turbulent Magnetic Diffusion}
 \bl
%%%%%%%%%%%%%%%%%%%%%%%%%%%%%%%%%%%%%%%%%%%%%%%%%%%%%%%%%%%%%%%%%%%
It was pointed out by K\"ohler (1973) that the 11-year period of solar cycles can be reproduced only if the turbulent magnetic diffusivity is reduced much below its usual mixing-length estimation $\eta_{_\mathrm{T}} \simeq 10^{13}$\,cm$^2$/s. A cyclic dynamo has to regenerate magnetic fields faster than diffusion destroys them. The cycle period should, therefore, be shorter than the diffusion time $t_\mathrm{d} = H^2/\eta_{_\mathrm{T}}$, where $H$ is the convection zone thickness ($H\simeq 200$\,Mm for the Sun). A small diffusivity, $\eta_{_\mathrm{T}} < 10^{12}$\,cm$^2$/s, is required to reproduce the 11-year period of the cycle.

The choice of the diffusivity value is not totally free, however. This is not only because the mixing-length profile of Fig.5 lies above $10^{13}$\,cm$^2$/s in the bulk of convection zone. Hydrodynamical models of global flows on the Sun are unstable with the eddy viscosity and thermal diffusivity below $10^{13}$\,cm$^2$/s (Tuominen et al. 1994). This means that the low diffusivities cannot be a correct parameterization of solar convection. As the eddy magnetic diffusion results from the same mixing as the eddy viscosity, it is not expected to be much smaller than the eddy viscosity. Direct numerical simulations of Yousef et al. (2003) show the turbulent magnetic Prandtl number of the order one. Miesch et al. (2012) found the low bound on the eddy diffusivity $\sim 10^{12}$\,cm$^2$/s by applying basic hydrodynamical relations to the differential rotation and meridional flow data. To sum-up, $10^{13}$\,cm$^2$/s is a plausible value for the eddy magnetic diffusivity in the bulk of the convection zone.

\begin{figure}[htb]
 \centerline{
 \includegraphics[width=12 cm]{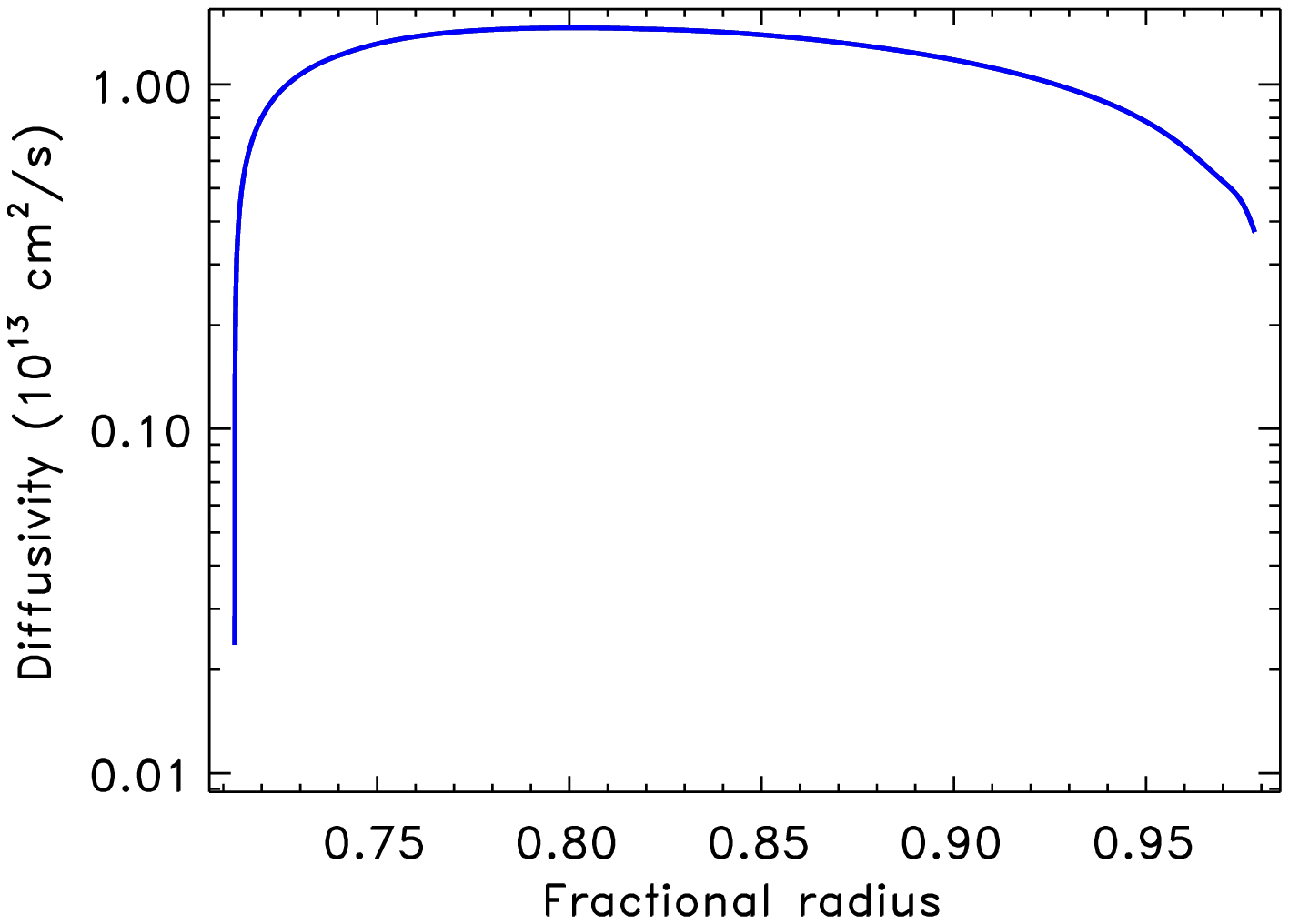}}
 \begin{description}
 \item{\small Fig.~5. Turbulent diffusivity profile in the solar convection zone estimated with the mixing-length relation $\eta_{_\mathrm{T}} = \ell_\mathrm{c} u_\mathrm{c}/3$ ($u_\mathrm{c}$ is the rms convective velocity; $\ell_\mathrm{c}$ is the mixing length) (Kitchatinov \& R\"udiger 2008).
    }
 \end{description}
\end{figure}

Fig.5 shows a sharp decline in the diffusivity with depth near the base of the convection zone. The problem of too short dynamo-cycles or too large diffusivity can be resolved by a pumping mechanism concentrating magnetic fields to the low diffusivity region near the base of the convection zone.
%%%%%%%%%%%%%%%%%%%%%%%%%%%%%%%%%%%%%%%%%%%%%%%%%%%%%%%%%%%%%%%%%%%
 \bll
 {\bf 3.1.2. Diamagnetic Pumping}
 \bl
%%%%%%%%%%%%%%%%%%%%%%%%%%%%%%%%%%%%%%%%%%%%%%%%%%%%%%%%%%%%%%%%%%%
The near-base region in the convection zone has long been recognized as a favorite place for the solar dynamo. Apart from the low diffusivity, this region can store strong magnetic fields against their buoyant rise. The field can be concentrated near the base by the diamagnetic pumping effect of inhomogeneous turbulence.

Inhomogeneously turbulent conducting fluids are known to act on large-scale fields as diamagnets to expel the fields from the regions of relatively high turbulence intensity. The fields are transported with the effective velocity
\begin{equation}
    \vec{U}_\mathrm{dia} = -\vec{\nabla}\eta_{_\mathrm{T}}/2
    \label{3}
\end{equation}
proportional to the gradient of magnetic diffusivity (Krause \& R\"adler 1980). Diamagnetic pumping has important consequences in solar dynamo models (R\"udiger \& Brandenburg 1995; Guerrero \& de Gouveia Dal Pino 2008). If a decrease in eddy diffusivity with depth across the base of convection zone is sufficiently large, magnetic fields in dynamo models are highly concentrated near the base (Kitchatinov \& Olemskoy 2012).

This concentration is also important for the generation of strong (kilogauss) toroidal fields. The relative magnitude of solar differential rotation is about 30\%. The Sun, therefore, has to perform about three rotations in order to produce a toroidal field equal in strength to the poloidal field from which the differential rotation winds it. In the 11 years of a solar cycle, a toroidal field about 40 times stronger can be wound. The large-scale polar field is about $1\div 2$\,G. If the field is distributed smoothly inside the Sun, differential rotation produces toroidal fields below 100\,G that, probably, does not suffice for sunspot formation. Poloidal fields should be much stronger somewhere inside the Sun compared to the surface for the differential rotation to wind kilogauss toroidal fields. The poloidal field concentration near the base of the convection zone can be provided by diamagnetic pumping.

Pumping also insures the dipolar equatorial symmetry of generated fields that dominates on the Sun (Stenflo 1988; Obridko et al. 2006). The downward pumping concentrates the toroidal field at the base of the convection zone. However, it does not influence the radial component of the poloidal field parallel to the pumping direction. The poloidal field, though also concentrated at the base, penetrates the convection zone up to its surface. Dipolar and quadrupolar global modes differ in the horizontal scale of the poloidal field, the scale being larger for dipolar modes. Relatively large turbulent diffusion in the bulk of the convection zone, therefore, affects stronger quadrupolar modes and results in preference for dipolar modes in dynamo models with a near-base concentration of magnetic fields (Chatterjee et al. 2004; Hotta \& Yokoyama 2010).
%%%%%%%%%%%%%%%%%%%%%%%%%%%%%%%%%%%%%%%%%%%%%%%%%%%%%%%%%%%%%%%%%%%
 \bll
 {\bf 3.1.3. Meridional Flow}
 \bl
%%%%%%%%%%%%%%%%%%%%%%%%%%%%%%%%%%%%%%%%%%%%%%%%%%%%%%%%%%%%%%%%%%%
Dynamo wave propagation along isorotational surfaces does not explain the observed
equatorward migration of sunspot activity. Wang et al. (1991) and Choudhuri et al. (1995)        proposed a very viable explanation of the equatorial drift by deep poleward meridional flow. The efficiency of magnetic field transport by meridional flow can be measured by the magnetic Reynolds number,
\begin{equation}
    R_\mathrm{m} = R_\odot V/\eta_{_\mathrm{T}} ,
    \label{4}
\end{equation}
where $V$ is the flow amplitude. Dynamo models with small diffusivity, $\eta_{_\mathrm{T}} < 10^{12}$\,cm$^2$/s, usually produce a too high polar concentration of the surface poloidal field. With the mixing-length value of $\eta_{_\mathrm{T}} \simeq 10^{13}$\,cm$^2$/s, the Prandtl number of Eq.(4) $P_\mathrm{m} \simeq 10$ is too small for the flow to produce a considerable effect in the bulk of the convection zone. The flow in the near-bottom low-diffusivity region (Fig.5) is however capable of producing the equatorial drift.

The poleward surface flow of the order of 10 \,m/s is observed on the Sun (Komm et al. 1993). Theoretical modeling predicts the internal flow to be concentrated in the boundary layers near the top (poleward flow) and bottom (equatorward) with a quite slow flow in between (Kitchatinov \& Olemskoy 2011b).   Helioseismology, however, does not support this picture. Multiple cells in radius (Zhao et al. 2013) or also in latitude (Schad et al. 2013) were detected. It is not clear whether it is possible to distinguish the global meridional flow from large-scale convection with a complicated spatial structure. Hazra et al. (2014) have shown that whatever structure the meridional flow has in the bulk of the convection zone, an advection dominated model results in a solar-like dynamo whenever the equatorward flow is present near the base of the convection zone.
%%%%%%%%%%%%%%%%%%%%%%%%%%%%%%%%%%%%%%%%%%%%%%%%%%%%%%%%%%%%%%%%%%%
 \bll
 {\bf 3.1.4. Catastrophic Quenching of the Alpha-Effect}
 \bl
%%%%%%%%%%%%%%%%%%%%%%%%%%%%%%%%%%%%%%%%%%%%%%%%%%%%%%%%%%%%%%%%%%%
Conservation of magnetic helicity presents a problem for the dynamo theory.  Briefly, the problem is as follows. Large-scale fields produced by the $\alpha$-effect are helical. As the magnetic helicity is conserved, small-scale fields attain their own helicity equal in absolute value and opposite in sign to the helicity of large-scale fields. Helical small-scale magnetic fields produce their own $\alpha$-effect, which is opposite in sign to the acting $\alpha$-effect of any origin. The total $\alpha$-effect strongly diminishes in the case of large magnetic Reynolds numbers (see Brandenburg \& Subramanian (2005) for more details).

However, catastrophic quenching fully applies only to a local formulation of the $\alpha$-effect. Only if the poloidal field is produced by the $\alpha$-effect at the same position that the toroidal field, from which it is produced, is located, is the resulting field necessarily helical. The Babcock-Leighton $\alpha$-effect discussed in Sect.2.3 is not local: a poloidal field near the surface is produced from a deep toroidal field. This type of $\alpha$-effect, therefore, is not necessarily subject to catastrophic quenching. However, non-locality of the $\alpha$-effect does not guarantee avoidance of catastrophic quenching. If a toroidal field is distributed smoothly with depth and has the same sign near the top and bottom, catastrophic quenching still applies (Brandenburg \& K\"apyl\"a 2007). This quenching does not occur if the regions of toroidal field concentration and poloidal field production by the (non-local) $\alpha$-effect are well separated in space (Kitchatinov \& Olemskoy 2011c).

A combination of diamagnetic pumping with the non-local alpha-effect of the Babcock-Leighton type seems to be promising for the solar dynamo modeling.
%%%%%%%%%%%%%%%%%%%%%%%%%%%%%%%%%%%%%%%%%%%%%%%%%%%%%%%%%%%%%%%%%%%
 \bll
 {\bf 3.2. Modeling the \lq\lq Cycle on Average''}
 \bl
%%%%%%%%%%%%%%%%%%%%%%%%%%%%%%%%%%%%%%%%%%%%%%%%%%%%%%%%%%%%%%%%%%%
Neglect of fluctuations in dynamo parameters results in strictly periodic magnetic cycles. The combination of diamagnetic pumping with non-local $\alpha$-effect helps to bring the simulated \lq\lq averaged'' cycles in close agreement with observations.

Figures 6 and 7 illustrate characteristic results of this type of models. The models produce global fields of dipolar parity. As discussed in Section 3.1.2, this is because of the magnetic field concentration near the base of the convection zone, where diffusion is relatively small. The toroidal field in Fig.6 is strongly concentrated at the bottom. The top row in this Figure augments the near-bottom region by expanding it in radius. The top dashed line in the upper row shows the radius just 4\% (in relative value) above the bottom. Otherwise, the toroidal field structure cannot be resolved by eye.  This concentration is due to diamagnetic pumping.  Though the model in Fig.6 belongs to the distributed type by its numerical design, it produces an interface dynamo.

The poloidal field of Fig.6 penetrates to the surface, but it is also concentrated at the base. The near-bottom poloidal field is about one hundred times stronger than on the top. Only due to this concentration, can the differential rotation in this model produce kilogauss toroidal fields.

\begin{figure}[htb]
 \centerline{
 \includegraphics[width=14 cm]{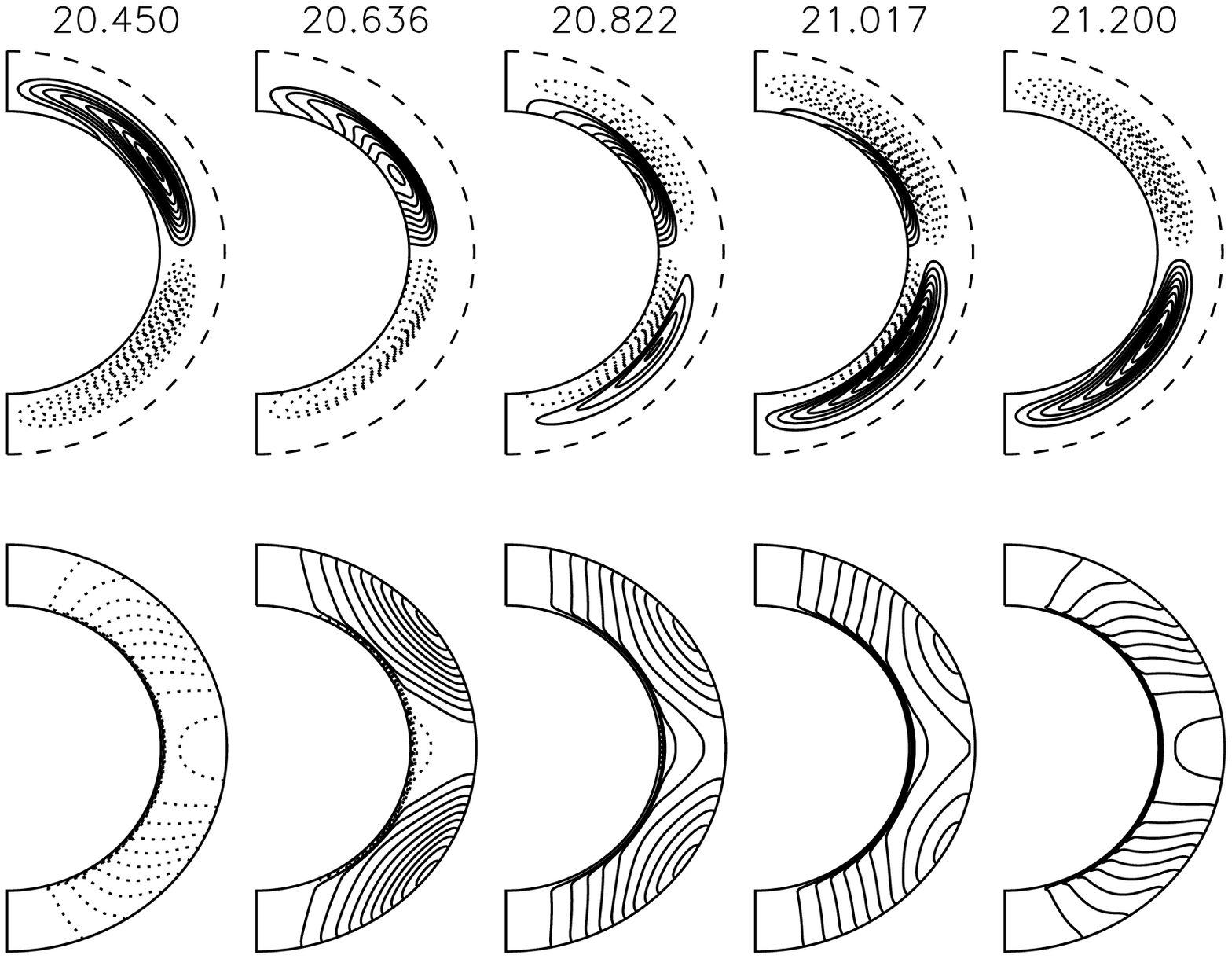}}
 \begin{description}
 \item{\small Fig.~6. Toroidal field isolines (top row) and poloidal field lines (bottom) in a dynamo model with diamagnetic pumping and non-local $\alpha$-effect. The numbers on the top show time in diffusive units $R^2_\odot/\eta_{_\mathrm{T}}$ (1 unit $\simeq$ 15 yrs). Full (dotted) lines show positive (negative) levels and clockwise (anti-clockwise) circulation of magnetic-field vectors. The dashed line in the top row shows the radius $r = 0.74R_\odot$ just $0.04R_\odot$ above the bottom boundary. After Kitchatinov \& Olemskoy (2012).
    }
 \end{description}
\end{figure}

\begin{figure}[htb]
 \centerline{
 \includegraphics[width=12 cm]{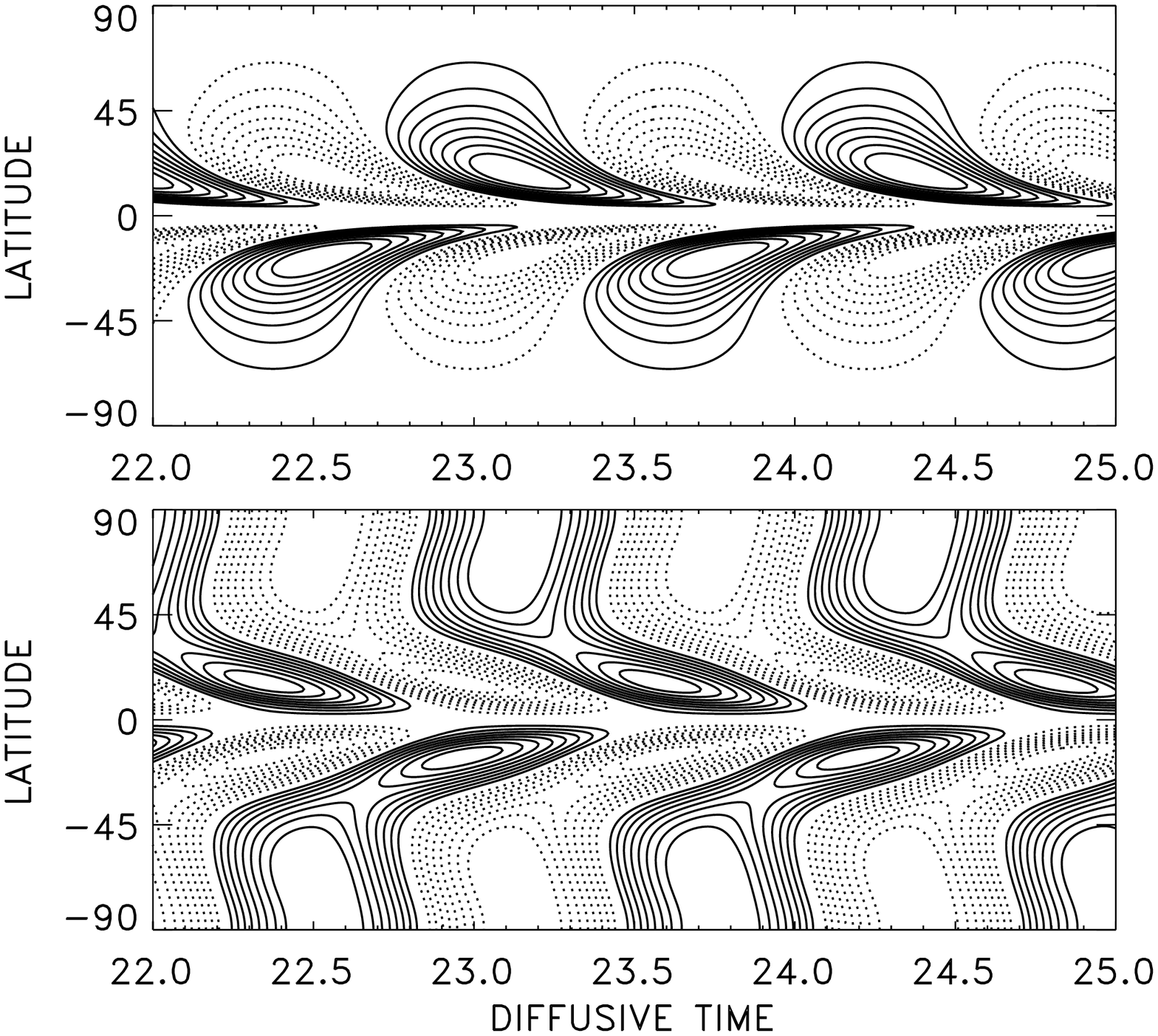}}
 \begin{description}
 \item{\small Fig.~7. Time-latitude diagrams of the near-bottom toroidal field (top) and the surface radial field after the same model as Fig.6. Equatorial drift of the toroidal field is produced by the near-bottom meridional flow.
    }
 \end{description}
\end{figure}

The butterfly diagram of Fig.7 shows equatorial drift and toroidal field maxima at low latitudes.  This is an effect of meridional flow towards the equator near the bottom. The flow in the bulk of the convection zone has a minor effect because of large magnetic diffusion (low $R_\mathrm{m}$ of Eq.(\ref{4})).

Though the model is generally close to observations, some disagreement can still be found. E.g., \lq\lq too early'' polar reversals prior to the cycle’s maxima can be noticed. This can probably be corrected by accounting for the decrease in magnetic diffusion towards the surface as can be seen in Fig.5. The magnetic cycles of this model are also \lq\lq too regular'' compared to the Sun. Observed cycles show variable durations and amplitudes. The irregularities are believed to result from fluctuations in dynamo parameters.
%%%%%%%%%%%%%%%%%%%%%%%%%%%%%%%%%%%%%%%%%%%%%%%%%%%%%%%%%%%%%%%%%%%
 \blll
 {\bf 3.3. Dynamos with Fluctuating Parameters}
 \bl
%%%%%%%%%%%%%%%%%%%%%%%%%%%%%%%%%%%%%%%%%%%%%%%%%%%%%%%%%%%%%%%%%%%
The dynamo theory explains irregularities in solar magnetic cycles by fluctuations in dynamo parameters (cf., e.g., Choudhuri 2013). Fluctuations in the $\alpha$-effect are the most relevant. The fluctuations can be estimated from sunspot statistics as discussed in Sect.2.3.

Figure 8 illustrates the irregular cycles computed in the dynamo model with fluctuating $\alpha$ (Olemskoy \& Kitchatinov 2013). The fluctuations vary irregularly with time and latitude. The model produces magnetic cycles of unequal durations. Irregular latitudinal variations in the $\alpha$-effect violate the equatorial symmetry of the generated fields so that the magnetic energy density is no longer symmetric about the equator. The $\alpha$-effect generates the poloidal magnetic field. Accordingly, the poloidal field diagram in Fig.8 is more irregular than the butterfly diagram. In particular, multiple reversals of the polar field can be seen in some of the magnetic cycles. As the field spreads over the convection zone and is pumped by the diamagnetic effect to the base of convection zone, its spatial distribution is smoothed by turbulent diffusion. The toroidal field produced by differential rotation at the base of convection zone is much more regular.

\begin{figure}[htb]
 \centerline{
 \includegraphics[width=12 cm]{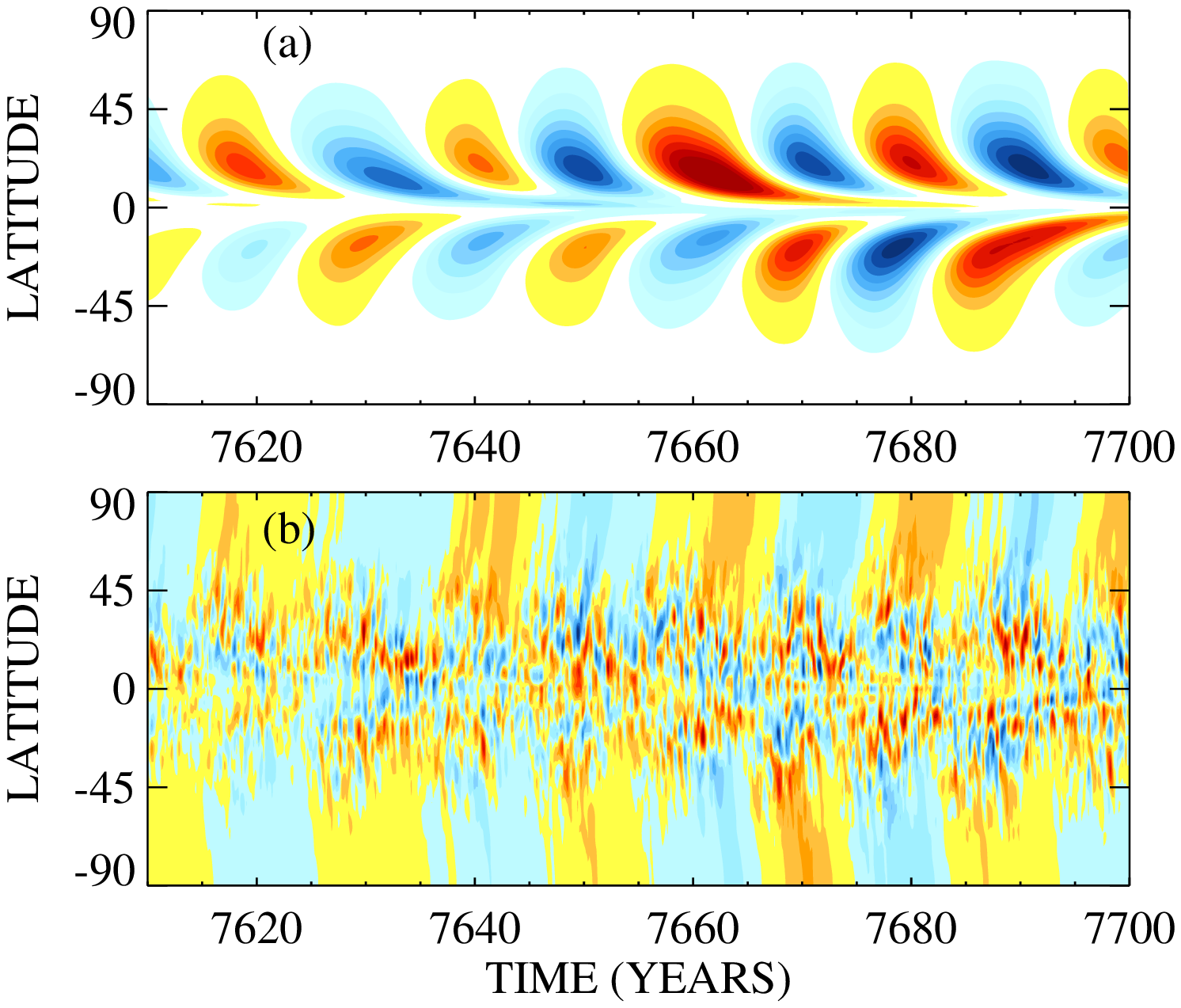}}
 \begin{description}
 \item{\small Fig.~8. Butterfly diagram of deep toroidal field (a) and time-latitude diagram of the surface radial field (b) in a dynamo model with fluctuating $\alpha$-effect. The amplitude of the fluctuations was specified after its estimation from sunspot statistics.
    }
 \end{description}
\end{figure}

Apart from fluctuations in the $\alpha$-effect, variations in the amplitude and structure of the meridional flow are significant. Fluctuations in the meridional flow are especially important in reproducing the Waldmaier relation – anti-correlation between the rise time of sunspot cycles and cycle amplitudes – with dynamo models (Karak \& Choudhuri 2011).

One of striking aspects of solar activity is that there were epochs in the past when activity levels were unusually low. A famous example is the Maunder minimum when few sunspots only were observed from the middle of 17th to the first quarter of 18th century. Radioisotope data on solar activity show that magnetic cycles did not disappear but were of unusually low strength during the Maunder minimum (Beer et al. 1998). The Maunder minimum is not a unique phenomenon in solar activity. Radoicarbon data show that there were about 27 grand minima in last 11000 years (Usoskin et al. 2007).

Epochs of magnetic cycles with strongly reduced amplitudes naturally result in dynamo models with fluctuating parameters (Choudhuri \& Karak 2012). Months and years long parametric fluctuations result in global variations in amplitude of dynamo-cycles on timescales of decades and centuries.  Memory time in fluctuating dynamos is short compared to these time scales (Yeates et al. 2008). Accordingly, the statistics of grand minima are close to the Poisson random process with the onset of each successive minimum statistically independent of preceding minima. Figure 9 compares distributions of durations of grand minima inferred from radiocarbon data with results of the dynamo model with fluctuating $\alpha$-effect. In both cases there are two groups of relatively short (30-90 yr) and relatively long (>110 yr) events.

\begin{figure}[htb]
 \centerline{
 \includegraphics[width=14 cm]{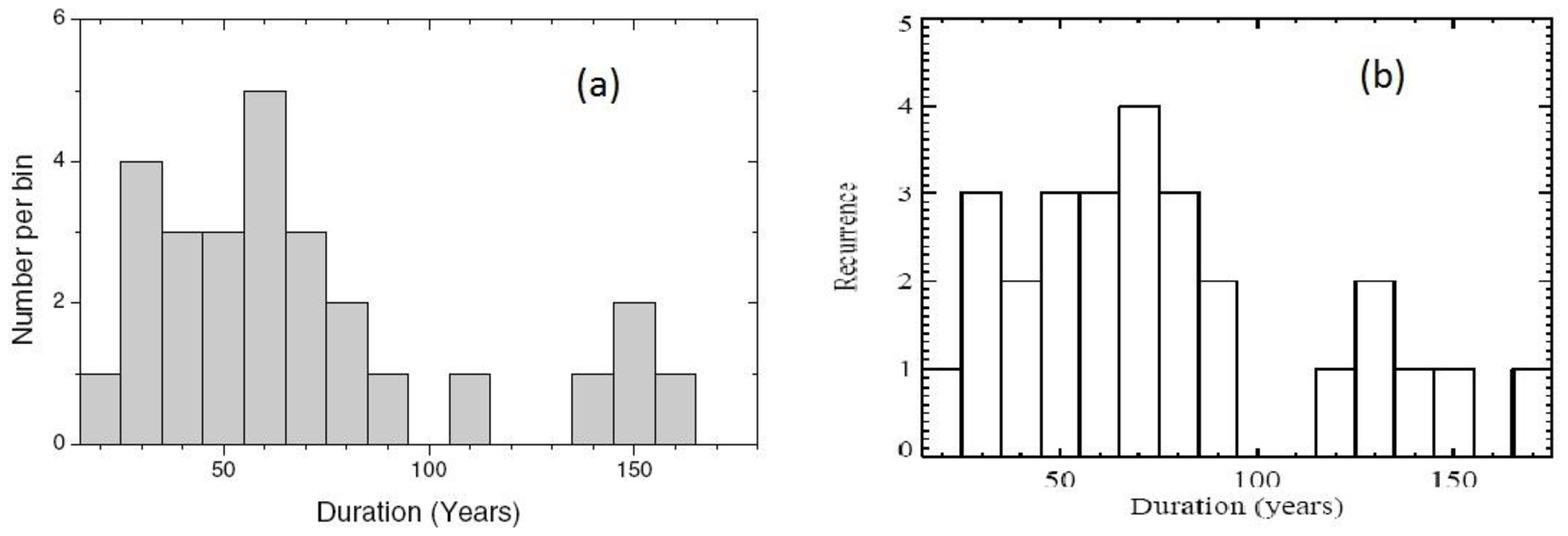}}
 \begin{description}
 \item{\small Fig.~9. Histograms of durations of grand minima from radiocarbon data (a) (Usoskin et al. 2007) and from the dynamo model with fluctuating $\alpha$-effect. The model used the fluctuation parameters estimated from sunspot statistics (Sect.2.3).
    }
 \end{description}
\end{figure}

It may be noted that grand minima and north-south asymmetry of solar activity may be related to each other. Almost all sunspots that appeared at the end of the Maunder minimum were in the southern solar hemisphere (Ribes \& Nesme-Ribes 1993). Theoretical models with fluctuations dependent not on time only but also on latitude also show this correlation (Usoskin et al. 2009; Olemskoy \& Kitchatinov 2013), which seems to have a clear physical meaning. Global fields in solar-type dynamo models are close to dipolar equatorial parity. Latitude-dependent fluctuations violate equatorial symmetry and transfer a part of the magnetic energy to fields of quadrupolar parity. The quadrupolar modes are subcritical for dynamo-excitation and decay. Violation of equatorial symmetry, therefore, tends to reduce magnetic energy. A large north-south asymmetry in solar activity can, therefore, be an indicator of transitions to grand minima. The relation between hemispheric asymmetry and activity level is, however, of statistical not deterministic nature.
%%%%%%%%%%%%%%%%%%%%%%%%%%%%%%%%%%%%%%%%%%%%%%%%%%%%%%%%%%%%%%%%%%%
 \bll
 {\bf 4. CONCLUSIONS AND PERSPECTIVES}
 \bl
%%%%%%%%%%%%%%%%%%%%%%%%%%%%%%%%%%%%%%%%%%%%%%%%%%%%%%%%%%%%%%%%%%%
The last decade has brought observational evidence for the operation of two basic effects of the dynamo theory on the Sun. It is possible to see from sunspot statistics that a non-local $\alpha$-effect of the Babcock-Leighton type takes part in the solar dynamo. The two basic effects of $\alpha\Omega$-dynamo do not, however, suffice to construct a realistic model. Only if the downward diamagnetic pumping and field advection near the base of the convection zone by equatorward flow are included, can general features of an “averaged” solar cycle be closely reproduced.

Observations favor operation of the Babcock-Leighton $\alpha$-effect on the Sun. This magnetic buoyancy-driven effect is, however, not as well understood as the \lq\lq standard'' $\alpha$-effect of inhomogeneous turbulence. In spite of recent progress in understanding magnetic buoyancy instability in rotating fluids (Chatterjee et al. 2011) and its operation in the presence of shear and downward pumping of a magnetic field (Barker et al. 2012), expressing the relevant $\alpha$-effect in terms of basic stellar parameters remains to be a perspective for the dynamo theory. Arbitrary prescriptions are, therefore, unavoidable in contemporary models of the solar dynamo. The arbitrariness also hinders extension of dynamo models to solar-type stars.

A very striking development was the reproduction of irregularities in solar activity cycles, in particular reproduction of grand minima, by dynamo models with fluctuating parameters. The physics of the grand minima phenomenon is still not clear, however. Characteristic times of fluctuations prescribed in different dynamo models for grand minima spread within two orders of magnitude. Relations between the fluctuations spectrum and parameters of simulated grand minima remains to be understood. The standard parametric resonance – including its stochastic version – does not exist in dynamos (Schmitt \& R\"udiger 1992; Moss \& Sokoloff 2013). Parametric excitation of irregularities in magnetic cycles hides still unknown physics.
%%%%%%%%%%%%%%%%%%%%%%%%%%%%%%%%%%%%%%%%%%%%%%%%%%%%%%%%%%%%%%%%%%%
 \bll
{\bf Acknowledgements.}
%%%%%%%%%%%%%%%%%%%%%%%%%%%%%%%%%%%%%%%%%%%%%%%%%%%%%%%%%%%%%%%%%%%

The author is thankful to the Russian Foundation for Basic Research for support (grant \n\  13-02-00277).
%%%%%%%%%%%%%%%%%%%%%%%%%%%%%%%%%%%%%%%%%%%%%%%%%%%%%%%%%%%%%%%%%%%

\bll

\centerline{\bf REFERENCES}
\begin{description}
\item Babcock, H.W.
    1961, {\sl ApJ} {\bf 133}, 572
\item Barker, A.J., Silvers, L.J., Proctor, M.R.E., \& Weiss, N.O.
    2012, {\sl MNRAS} {\bf 424}, 115
\item Beer, J., Tobias, S., \& Weis, N.
    1998, {\sl Solar Phys.} {\bf 181}, 237
\item Benz, A.O.
    2008, {\sl Living Reviews in Solar Physics} {\bf 5}, 1
\item Brandenburg, A., \& K\"apyl\"a, P.
    2007, {\sl New Journal of Physics} {\bf 9}, 305
\item Brandenburg, A., \& Subramanian, K.
    2005, {\sl Physics Reports} {\bf 417}, 1
\item Chatterjee, P., Nandy, D., \& Choudhuri, A.R.
    2004, {\sl A\&A} {\bf 427}, 1019
\item Chatterjee, P., Mitra, D., Rheinhardt, M., \& Brandenburg, A.
    2011, {\sl A\&A} {\bf 534}, A46
\item Choudhuri, A.R.
    1992, {\sl A\&A} {\bf 253}, 277
\item Choudhuri, A.R.
    2013, arXiv:1312.3408
\item Choudhuri, A.R., Sch\"ussler, M., \& Dikpati, M.
    1995, {\sl A\&A} {\bf 303}, L29
\item Erofeev, D.V.
    2004, An Observational Evidence for the Babcock-Leighton Dynamo Scenario, in: {\sl Proc. IAU Symp. 223 Multi-Wavelength Investigations of Solar Activity.} A.V. Stepanov, E.E  Benevolenskaya \& A.G. Kosovichev (eds.), Cambridge Univ. Press,
    pp.97-98
\item Guerrero, G., \& de Gouveia Dal Pino, E.M.
    2008, {\sl A\&A} {\bf 485}, 267
\item Hotta, H., \& Yokoyama, T.
    2010, {\sl ApJ} {\bf 714}, L308
\item Hazra, G., Karak, B.B., \& Choudhuri, A.R.
    2014, {\sl ApJ} {\bf 782}, 93
\item Howard, R.F.
    1996, {\sl ARA\&A} {\bf 34}, 75
\item Howard, R., Gilman, P.I., \& Gilman, P.A.
    1984, {\sl ApJ} {\bf 283}, 373
\item Howard, R.F., Gupta, S.S., \& Sivaraman, K.R.
    1999, {\sl Solar Phys.} {\bf 186}, 25
\item Howard, R., \& LaBonte, B.J.
    1980, {\sl ApJ} {\bf 239}, L33
\item Hoyng, P.
    1988, {\sl ApJ} {\bf 332}, 857
\item Ivanova, T.S., \& Ruzmaikin, A.A.
    1976, {\sl SvA} {\bf 20}, 227
\item Jiang, J., Chatterjee, P., \& Choudhuri, A.R.
    2007, {\sl MNRAS} {\bf 381}, 1527
\item Karak, B.B., \& Choudhuri, A.R.
    2011, {\sl MNRAS} {\bf 410}, 1503
\item Karak, B.B., \& Choudhuri, A.R.
    2012, {\sl Phys. Rev. Lett.} {\bf 109}, 171103
\item Kitchatinov, L.L., \& Olemskoy, S.V.
    2011a, {\sl Astron. Lett.} {\bf 37}, 656
\item Kitchatinov, L.L., \& Olemskoy, S.V.
    2011b, {\sl MNRAS} {\bf 411}, 1059
\item Kitchatinov, L.L., \& Olemskoy, S.V.
    2011c, {\sl Astron. Nachr.} {\bf 332}, 496
\item Kitchatinov, L.L., \& Olemskoy, S.V.
    2012, {\sl Solar Phys.} {\bf 276}, 3
\item Kitchatinov, L.L., \& R\"udiger, G.
    2008, {\sl Astron. Nachr.} {\bf 329}, 372
\item K\"ohler, H.
    1973, {\sl A\&A} {\bf 25}, 467
\item Komm, R.W., Howard, R.F., \& Harvey, J.W.
    1993, {\sl Solar Phys.} {\bf 147}, 207
\item Krause, F., \& R\"adler, K.-H.
    1980, {\sl Mean-Field Magnetohydrodynamics and Dynamo Theory},
    Akademie-Verlag, Berlin
\item Leighton, R.B.
    1969, {\sl ApJ} {\bf 156}, 1
\item Maehara, H., Shibayama, T., Shota, N. et al.
    2012, {\sl Nature} {\bf 485}, 478
\item Makarov, V.I., \& Tlatov, A.G.
    2000, {\sl Astron. Rep.} {\bf 44}, 759
\item Makarov, V.I., Tlatov, A.G., Callebaut, D.K. et al.
    2001, {\sl Solar Phys.} {\bf 198}, 409
\item Miesch, M.S., Featherstone, N.A., Rempel, M., \& Trampedach, R.
    2012, {\sl ApJ} {\bf 757}, 128
\item Moss, D., \& Sokoloff, D.
    2013, {\sl A\&A} {\bf 553}, A37
\item Moss, D., Sokoloff, D., Usoskin, I., \& Tutubalin, V.
    2008, {\sl Solar Phys.} {\bf 250}, 221
\item Obridko, V.N., Sokoloff, D.D., Kuzanyan, K.M., Shelting, B.D., \& Zakharov, V.G.
    2006, {\sl MNRAS} {\bf 365}, 827
\item Olemskoy, S.V., Choudhuri, A.R., \& Kitchatinov, L.L.
    2013, {\sl Astron. Rep.} {\bf 57}, 458
\item Olemskoy, S.V., \& Kitchatinov, L.L.
    2013, {\sl ApJ} {\bf 777}, 71
\item Ossendriver, A.J.H., Hoyng, P., \& Schmitt, D.
    1996, {\sl A\&A} {\bf 313}, 938
\item Parker, E.N.
    1955, {\sl ApJ} {\bf 122}, 293
\item Ribes, J.C., \& Nesme-Ribes, E.
    1993, {\sl A\&A} {\bf 276}, 549
\item R\"udiger, G., \& Brandenburg, A.
    1995, {\sl A\&A} {\bf 296}, 557
\item Schad, A., Timmer, J., \& Roth, M.
    2013, {\sl ApJL} {\bf 778}, L38
\item Schatten, K.H., Scherrer, P.H., Svalgaard, L., \& Wilcox, J.M.
    1978, {\sl Geophys. Res. Lett.} {\bf 5}, 411
\item Schmitt, D., \& R\"udiger, G.
    1992, {\sl A\&A} {\bf 264}, 319
\item Shibata K., Isobe H., Hillier A. et al.
    2013, {\sl Publ. Astron. Soc. Japan} {\bf 65}, 49
\item Steenbeck, M., \& Krause, F.
    1969, {\sl Astron. Nachr.} {\bf 291}, 49
\item Steenbeck, M., Krause, F., \& R\"adler, K.-H.
    1971, A Calculation of the Mean Electromotive Force in an Electrically Conducting Fluid in Turbulent Motion, Under the Influence of Coriolis Forces, in: {\sl The Turbulent Dynamo (A translation of series of papers by F. Krause, K.-H. R\"adler, and M. Steenbeck), P.H. Roberts \& M. Stix (eds.), NCAR, Boulder, Colorado},  pp.29-47
\item Stenflo, J.O.
    1988, {\sl  Ap\&SS} {\bf 144}, 321
\item Stix, M.
    1976, {\sl A\&} {\bf 47}, 243
\item Svalgaard, L., Cliver, E.W., \& Kamide, Y.
    2005, {\sl Geophys. Res. Lett.} {\bf 32}, L01104
\item Tuominen, I., Brandenburg, A., Moss, D., \& Rieutord, M.
    1994, {\sl A\&A} {\bf 284}, 259
\item Usoskin, I.G., Solanki, S.K., \& Kovaltsov, G.A.
    2007, {\sl A\&A} {\bf 471}, 301
\item Usoskin, I.G.,  Sokoloff, D., \& Moss, D.
    2009, {\sl Solar Phys.} {\bf 254}, 345
\item Vorontsov, S.V., Christensen-Dalsgaard, J., Schou, J. et al.
    2002, {\sl Scince} {\bf 296}, 101
\item Wang, Y.-M., Sheeley, N.R. Jr., \& Nash, A.G.
    1991, {\sl ApJ} {\bf 383}, 431
\item Yeates, A.R., Nandy, D., \& Mackay, D.H.
    2008, {\sl ApJ} {\bf 673}, 544
\item Yousef, T.A., Brandenburg, A., \& R\"udiger, G.
    2003, {\sl A\&A} {\bf 411}, 321
\item Zhao, J., Bogart, R.S., Kosovichev, A.G., Duvall, T.L. Jr., \& Hartlep, T.
    2013, {\sl ApJL} {\bf 774}, L29
\end{description}
\end{document}